# SOPC Co-Design Platform for UWB System in Wireless Sensor Network Context

Daniela Dragomirescu
LAAS-CNRS
University of Toulouse
7, Av du Colonel Roche
31077 Toulouse cedex 4, France
daniela@lass.fr

Aubin Lecointre
LAAS-CNRS
University of Toulouse
7, Av du Colonel Roche
31077 Toulouse cedex 4, France
alecoint@laas.fr

Robert Plana
LAAS-CNRS
University of Toulouse
7, Av du Colonel Roche
31077 Toulouse cedex 4, France
plana@laas.fr

*Abstract* – This paper presents our approach of the radio interface problematic for Wireless Sensor Network. We introduce the WSN context and constraints associated. We propose an IR-UWB solution and illustrate why it could be a viable solution for WSN. A high level modelling and simulation platform for IR-UWB radio interface is proposed on Matlab. It allows us to determine according to BER versus Eb/N0 criteria and the WSN constraints what kind of design is more adequate. Moreover, a co-design co-simulation platform Matlab VHDL is proposed here. Using this platform we designed IR-UWB transceiver having reconfigurable capabilities, such as data rate reconfiguration, time hopping code, spectrum occupation and radio range reconfiguration.

## I. Introduction

We lead our study in the context of wireless sensor network (WSN). Our goal is to propose a radio interface for this kind of network. We define WSN as systems having very large number of nodes on a small area. WSN is a WPAN-like concept (Wireless Personal Area Network). There are a lot of kinds of applications for this variety of networks; such as: monitoring, military applications, house automation, civil safety applications, etc. By considering these applications, we could deduce easily that there are some intrinsic constraints for WSN, which are: low cost, low power, simplicity and tiny nodes. Indeed, without theses characteristics none networks could be a viable WSN. Thus all along this paper we keep in mind this context in order to solve locks of WSN radio interface

We could note that there is an important diversity of WSN applications; consequently it's difficult to propose a radio interface for WSN. There is two way as response for WSN diversity applications. The first one consists in dividing WSN applications in few main important categories and proposes an optimized radio interface for each one. The second way is a kind of absolute solution: the goal is to implement a reconfigurable radio interface. This latter will be able to answer to the different needs and constraints, whatever the WSN application considered. This second way is the most innovative way, and it proposes the implementation of reconfigurablity concept inspired from software-defined radio [1]. We will follow this concept throughout this paper.

To achieve this goal, i.e. design a WSN radio interface, we will deal with the problematic in a global way. That is to say, that we will study this radio interface problem from the high level to the hardware level. Here, we propose also to address the topic at different stages, such as emitter, RF channel and receiver. We will introduce our co-design co-modelling and co-simulation platform. Each choice will be exposed and detailed during this paper (such as wireless technique, hardware implementation …). Thus, we will explain the IR-UWB principle and why we selected it as our radio technology [2]. Then we will show the concept of our Matlab IR-UWB platform which aims is to model, simulate and validate the IR-UWB radio interface (PHY layer). After this indispensable and precondition phase of modelling and simulation, we would be able to implement our radio interface on a FPGA, Field Programmable Gate Array or on ASIC (Application Specific Integrated Circuit).

This paper is laid out as follows: Section II presents the technology used for address the WSN constraints and goals: Impulse Radio Ultra WideBand (IR-UWB). Section III describes the high level modelling and simulation platform developed on Matlab. Hardware implementation and simulation on FPGA/ASIC will be introduced in the section IV, before conclusion in the section V.

## II. UWB for Wireless Sensor Network

Considering the four specific WSN constraints, cost, power, size, simplicity, we chose to use UWB. Indeed this technology, since the FCC (Federal Communications Commission) allocated the unlicensed 3,1 – 10,6 GHz band for UWB in 2002, seems to be adapted to WSN context. The Federal Communications Commission (FCC) defines a radio system to be an UWB system if the -10 dB bandwidth of the signal is at least 500 MHz or the fractional bandwidth is greater than 20% [2], with the fractional band:

$$B_f = 2 \times \frac{f_H - f_L}{f_H + f_L}$$

Impulse Radio UWB (IR-UWB) is a very promising technology for the WSN applications. Let us quote these advantages: 7,5 GHz of free spectrum which could permit to reach high data rate, extremely low transmission energy, extremely difficult to intercept, multi-path immunity, low cost (mostly digital architecture), "Moore's Law Radio" (performances, size, data rate, cost follow Moore's Law), simple CMOS transmitter at very low power [2][3]. Among the various families within UWB, we focus on family IR-UWB, Impulse Radio UWB which is appropriate for our context of application: wireless sensor network. It consist of send very short pulses (< 1ns) on the channel.

To transmit information by means of UWB pulse, we must use pulse modulation. The IR-UWB modulations are: PPM (Pulse Position Modulation), PAM (Pulse Amplitude Modulation), OOK (On Off Keying), BiPhase Modulation, PSM (Pulse Shape Modulation) [2]. For the radio interface, we have to choose among these modulations (figure 1):

- PPM differs bits by a time shift.
- PAM represents binary one and binary zero by pulses with distinct amplitude ($A_1 \neq A_0$).
- OOK is a special case of PAM. Binary one are represented with a pulse amplitude of $A_1 = A$, when binary zero use $A_0 = 0$.
- PSM sends distinct pulse waveform to represent binary zero and binary one.
- BiPhase could be considered as a special case of PAM when $A_1 = A$ and $A_0 = -A$. In this case, we observed a shifting of 180° in the two pulses.

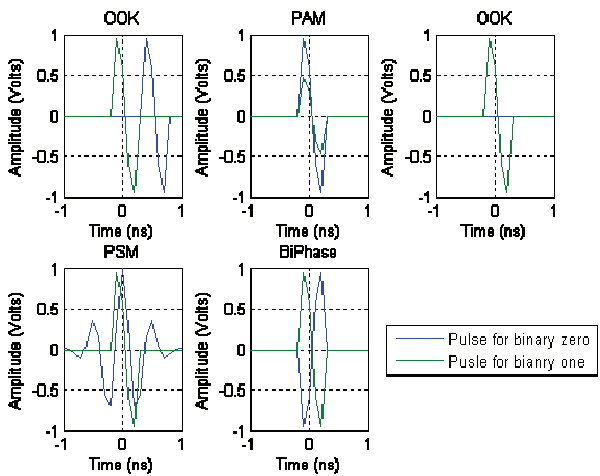

Figure 1. IR-UWB Pulse Modulations

If we visualize the power spectral densities of these modulation, we could observe narrow line spectral components due to the repetition of pulse at a given rate. We use Time Hopping (TH) for suppressing this drawback. Line spectral components are undesirable regarding to the power FCC regulation [3][4]. TH permits to smooth the spectrum, adds the multi-user capability to the transmission [5], decreases the probability of collision (between pulses from distinct users) [6], and confers to the signal a low probability of interception and detection by having a noise-like spectrum. This spread spectrum technique consists in dividing the channel in successive frames. Each frame is compound of time slots (or chips) which contain the modulated pulse according the code scheme. The user repartition on different time slots is carry out by the use of Time Hopping code (TH-Code) associated with each user. TH is like a dynamic TDMA. The figure 2 exposes the Time Hopping principle in the case of two users.

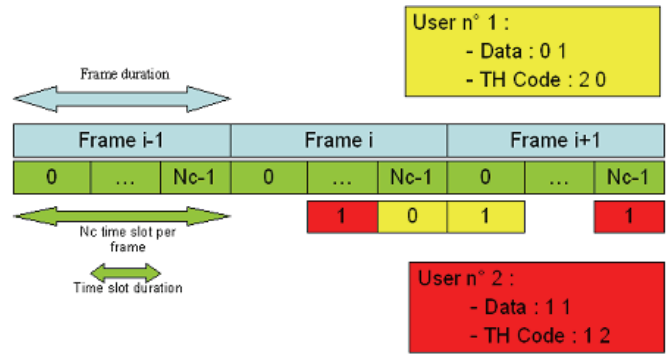

Figure 2. Illustration of Time Hopping with two users

In this part we have introduce the innovative radio technology: IR-UWB, which seems to be efficient for WSN. Now it should be interesting to determine what kind of architecture (emitter and receiver) will be better (Bit Error Rate performance BER) considering distinct channels. It will be the topic of the next part.

III. HIGH LEVEL MODELLING AND SIMULATION PLATFORM

We use Matlab for developing our IR-UWB platform. It takes into account the whole IR-UWB link, from multi-users emitters to receivers, including also radio channel. Figure 3 illustrates this high level platform.

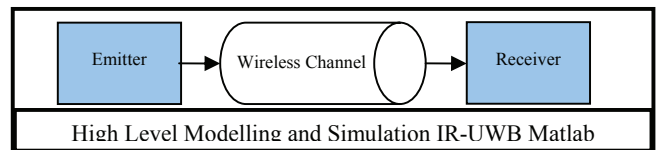

Figure 3. High Level Platform

One of the most important advantages of this platform is its capability to deal with a lot of configurations, thanks to its parametric design. Indeed, we are able to simulate, evaluate the performance of any kind combination of the following platform parameters:

- Parameters at the emission:
  o UWB pulse properties (amplitude, duration, shape), IR-UWB modulation (PPM, PSM, OOK, BPAM, BiPhase), Time Hopping parameters (chip duration, number of chip per frame), number of users, mobility speed of the emitter.

- Kind of channel:
  o Perfect channel, AWGN (Additive White Gaussian Noise) channel, IEEE 802.15.4a

- UWB 2-10GHz channel, UWB 60GHz channel (IEEE 802.15.3c)
    - Propagation delay, path loss, …
- Receiver parameters:
    - Coherent/non-coherent, simple/double correlation, distance estimation, IR-UWB demodulation (PPM, PSM, OOK, BPAM, BiPhase), synchronization mechanism, kind of channel estimation (at 60 GHz, 3-10 GHz and under 1 GHz).

This platform allows us on the one hand to implement, and understand concretely a full IR-UWB link with localization capability over a realistic channel (IEEE 802.15.4a [7][8]). On the other hand, it permits us to estimate the performance of the different proposed radio interface, according the $E_b/N_0$ (energy per bit to noise power spectral density ratio) versus BER (Bit Error Rate) criteria. It permits to classify the performances of the modulation and the receiver technique. This platform is vital in order to determine which solution in term of design architecture is the most adequate for our WSN context. At this point we have information about system level performances. Further, the hardware level will inform us about expected performance regarding physical implementation (size, power consumption, frequency, …). Thus we will able to choose a trade off between system and hardware level performances.

For example, let us compare two solutions according to the BER Eb/N0 criteria and the four WSN constraints (size, cost, simplicity, power). We compare the TH-PPM IR-UWB versus TH-OOK IR-UWB.

For TH-OOK, we propose a non-coherent detection energy based receiver described in figure 4 [9]. This receiver is less expensive, simpler, less greedy in power consumption, and it has smaller overall dimensions than the TH-PPM receiver described in figure 5.

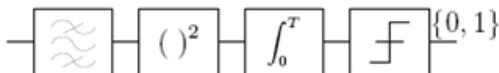

Figure 4. Non coherent OOK receiver

The TH-PPM coherent receiver is based on the correlation, with a template waveform, principle [2].

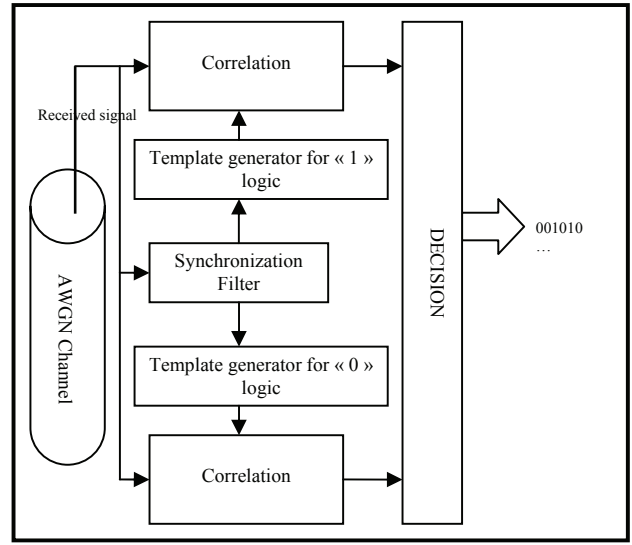

Figure 5. TH-PPM coherent receiver

A synchronization bloc is also necessary, in order to provide a synchronous correlation between the received pulse and the template waveform. This function adds complexity to the receiver. This synchronization is difficult, because of the pulse duration (< 1 ns) and should be the most precise possible, because performances depend to the synchronization accuracy. Indeed, the synchronization will impact the BER performance for a coherent receiver (e.g. TH-PPM receiver). With this kind of receiver, while the synchronization is not done precisely, the BER will be bad and tends toward 0.5. That's why a precise synchronization is required for achieving good BER performance with such kind of coherent receiver.

Table I summarizes the comportment of our two considered solutions at the system level. While figure 6 exposes the BER versus Eb/N0 criteria on an UWB 3-10 GHz channel (IEEE 802.15.4a) for TH-PPM, TH-OOK and three classical narrow band techniques [10]. We will compare these results at system level with results obtained at hardware level in a further part of the paper.

TABLE I. COMPARATIVE ANALYSIS OF IR-UWB ARCHITECTURES

| Classification | WSN Constraints | | | | |
|---|---|---|---|---|---|
| IR-UWB for WSN | Power | Cost | Simplicity | Size | BER vs SNR |
| TH-PPM | 2 | 2 | 2 | 2 | 1 |
| TH-OOK | 1 | 1 | 1 | 1 | 2 |

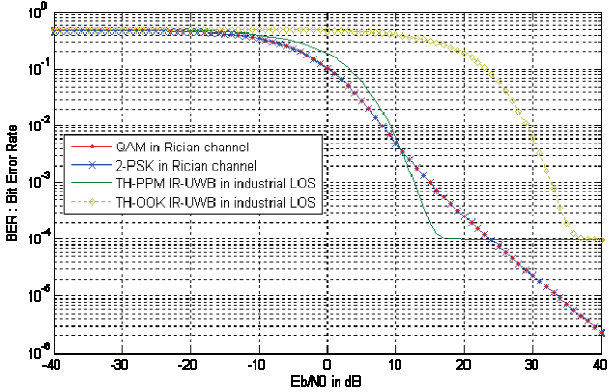

Figure 6. IR-UWB techniques in UWB channel: residential LOS, versus continuous carrier wave technique in Rician channel (K=4), according to the BER versus Eb/N0 criteria.

Figure 6 proves that IR-UWB techniques offer approximately the same BER performance, than classical narrow band solutions (PSK, and QAM). This comparison depends also in the similitude of the two channels. It illustrates also the fact that TH-PPM is better of about 15 dB than TH-OOK. While table II shows us that TH-OOK is more adequate than TH-PPM in order to deal with the four WSN constraints. Thus we can say there is a balance between performances and WSN constraints. We will examine this trade off also at hardware level.

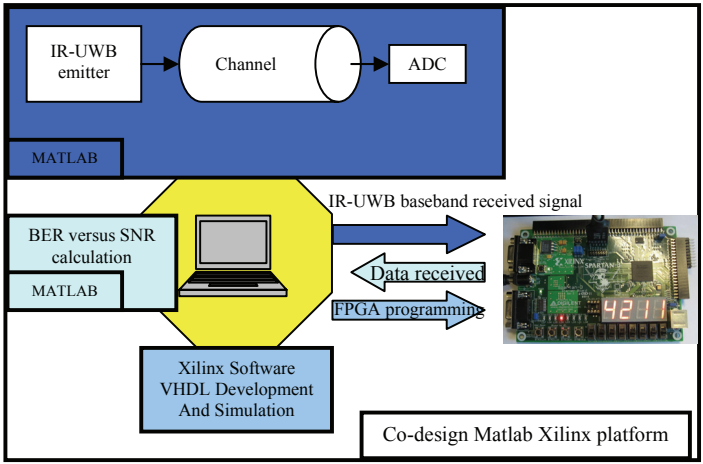

Figure 7. Co-simulation and co-performances analysis Matlab Xilinx Platform.

## IV. LOW LEVEL IMPLEMENTATION AND SIMULATION ON FPGA AND ASIC

Thanks to our IR-UWB Matlab platform (fig. 7) we have modelling and validate IR-UWB solutions for WSN radio interface. In this part we will implement them on Xilinx Spartan III FPGA [11], because it is a cheaper and an optimized signal processing solution. Nevertheless, Virtex 5 implementation could be also used and ASIC implementation is proposed for power consumption information. We use the co-simulation co-design and co-performances analysis Matlab Xilinx platform described figure 7.

We re-use the two previous configurations: TH-OOK and TH-PPM in order to study them at hardware level.

This platform allows us to compare, classify a lot of distinct architectures, here we will start with a comparison of TH-OOK and TH-PPM transceiver as illustrated in table II by using an ASIC target. The platform uses power consumption size and frequency criteria and BER criteria, for achieving an interesting classification of ours transceivers architecture in WSN context.

TABLE II. COMPARATIVE ANALYSIS OF TH-PPM AND TH-OOK IR-UWB ARCHITECTURES

| Version | ASIC Design Kit AMS 3.60 at 0.35µm | | |
|---|---|---|---|
| | Maximum Frequency | Size | Power Consumption |
| Static TH-PPM Emitter | 269,5 MHz | 47666 | 3,2mW @ 100MHz |
| | | | 6,5mW @ 200MHz |
| | 529,1 MHz | 77500 | 22,5mW @ 500MHz |
| Static TH-PPM Receiver | 173,6 MHz | 49858 | 3,5mW @ 100MHZ |
| | 216,4 MHz | 50367 | 7,1mW @ 200MHZ |
| | 357,1 MHz | 64483 | 11,8 mW @ 285MHz |
| Static TH-OOK Emitter | 280,8 MHz | 42857 | 2,9mW @ 100MHz |
| | | | 5,8mW @ 200MHz |
| | 606 MHz | 71302 | 19,4mW @ 500MHz |
| Static TH-OOK Receiver | 221,7 MHz | 33625 | 2,5mW @ 100MHZ |
| | | | 5mW @ 200MHz |
| | 568,1 MHz | 50517 | 15 mW @ 500MHz |

Table II shows us that OOK transceiver is better regarding WSN criteria (size, power consumption …) than TH-PPM solution at both sides: emitter and receiver. Indeed, we could see that power consumption is lesser with OOK than with PPM. In addition the size of the OOK circuits are also smaller than with PPM transceiver. Concerning the maximum frequency of the circuit, i.e. the operating frequency, OOK solution permits to achieve better performances, since with these kinds of time domain techniques, TH IR-UWB, the data rate is function of the maximum operating frequency. As a results OOK architectures proposes better performances while offering a better behaviour regarding WSN context; nevertheless, the BER performance is better in the PPM case, as illustrated with system level modelling. Consequently, there is a trade off de set up in function of the considered WSN application.

Table II compares OOK and PPM transceiver architecture at hardware level without the synchronization stage required in coherent approach such as PPM. We have chosen to separate the modelling of the synchronization because it's a problem itself in IR-UWB. Synchronization technique is a key point in IR-UWB solution due to the very small duration of the pulse used. It's in this case, typically, that our co-design platform could be used. Indeed, there are a lot of possible synchronization techniques for a coherent PPM, a such platform allows to compare them both at hardware and system level. Table III summarizes this kind of comparison at hardware level.

TABLE III. AN EXAMPLE OF A COMPARATIVE ANALYSIS OF SYNCHORNIZATION TECHNIQUES IN IR-UWB

| Version | ASIC Design Kit AMS 3.60 at 0.35µm | | |
|---|---|---|---|
| | Maximum Frequency | Size | Power Consumption |
| Synchronization : matched filter v1 | 84,1 MHz | 798493 | 18,3mW @ 83,3MHz |
| Synchronization : matched filter v2 | 101,5 MHz | 507958 | 14,7mW @ 100MHz |
| | 126,9 MHz | 533700 | 19,9mW @ 125MHz |

We could see that the two distinct version of the proposed synchronization offer distinct performances regarding size and power consumption. Let's us note there are a lot of existing synchronization techniques in IR-UWB. Here, these two one are only used as illustration. We have to consider carefully this synchronization problematic since it seems, thanks to table III, to be expensive regarding WSN constraints. In addition, in our OOK-PPM comparison example, the synchronization cost will increase the gap between the two solutions.

Another example of the use of this co-design platform could be used for determining the interest of the reconfigurability in WSN context. [12] Reconfigurability proposed here permits to change the data rate, the spectrum occupation, the radio range and the TH-code of the transceiver without requiring a FPGA re-programming for example. Table IV illustrates the cost of reconfigurability regarding WSN criteria.

TABLE IV. EXPRESSION OF THE COST OF RECONFIGURABILITY IN WSN CONTEXT WITH ASIC IMPLEMENTATION.

| Version | ASIC Design Kit AMS 3.60 at 0.35µm | | |
|---|---|---|---|
| | Maximum Frequency | Size | Power Consumption |
| Static TH-PPM Emitter | 269,5 MHz | 47666 | 3,2mW @ 100MHz |
| | | | 6,5mW @ 200MHz |
| | 529,1 MHz | 77500 | 22,5mW @ 500MHz |
| Reconfigurable TH-PPM Emitter | 222,7 MHz | 45032 | 3,5mW @ 100MHz |
| | | | 7,1mW @ 200MHz |
| | 346 MHz | 49623 | 13,8mW @ 333MHz |

Table IV demonstrates the cost of reconfigurability at hardware level in WSN context. Indeed, the reconfigurability implies a decrease of the maximum achievable frequency, as a result a decrease of the maximum achievable data rate. Concerning WSN criteria, for example, the power consumption, the reconfigurability proposed here, imply an increase of the consumption for supporting this new functionality. Table V makes the same estimation of the cost of reconfigurability by using a FPGA Spartan 3 as target, which is a capability of our co-design platform (ASIC and FPGA target). With table V we could expose that on FPGA, the cost of reconfigurability is the same than on ASIC: a decrease of the maximum frequency and thus a smaller achievable data rate.

TABLE V. EXPRESSION OF THE COST OF RECONFIGURABILITY IN WSN CONTEXT WITH FPGA IMPLEMENTATION.

| Version | Synplify FPGA Spartan 3 : Place & Route | |
|---|---|---|
| | Maximum Frequency | Size (gates) |
| Static TH-PPM Receiver | 150,784 MHz | 1846 + 912 |
| Reconfigurable TH-PPM Receiver | 112,790 MHz | 20669 + 2592 |

The design is thought in a modular concept way. We have determined some vital function that we have implement in small blocs. Then it is enough to connect appropriates blocs in order to set up the desired system.

In conclusion, our co-design and co-simulation platform allows us to design the outline of the future PHY layer of the reconfigurable radio interface for WSN. We have shown that TH-PPM architecture propose better BER performance than TH-OOK system at the cost of less respect to the WSN constraints, especially if we consider the synchronization issue. We will also note that reconfigurability implies an increase of the power consumption and a decrease of the maximum frequency (and consequently the possible data rate). Consequently we notice a balance, exposed at the high level modeling and hardware level, i.e. the better BER performances are, the lesser WSN constraints are optimized. Here the OOK solution is better regarding WSN constraints, while PPM solution is better for BER performance criteria.

At last, it's important to note, that the way used here thanks to our co-design platform is replicable for comparing and classifying a large range of WSN architecture. Here we have shown results concerning OOK-PPM comparison, the impact of reconfigurability and the synchronization issue in IR-UWB, but it's also possible to work for example on BPAM techniques or PSM techniques or DS-UWB (Direct Sequence UWB), …

V. CONCLUSION

In this paper we introduce our work on radio interface for wireless sensor network. In a first time, we have exposed the specificities of the WSN context and their implications on the desired radio interface. Let us quote for example the low power aspect, the need of simplicity, small size circuit, and the low cost constraint. Then we have shown why we have proposed to use IR-UWB system in our WSN radio interface. Thanks to our high level modeling and simulating IR-UWB Matlab platform we have proved that TH-PPM and TH-OOK are more adequate and offer under condition the same performances (BER versus Eb/N0 criteria) as classical narrow band techniques for WSN applications. Finally, we have also implement our IR-UWB receivers on FPGA and ASIC by means of a modular concept way. We have compared them regarding WSN constraints (cost, size, low power, simplicity) and BER versus Eb/N0 performance. This study leads us to a balance between performance and the respect of WSN constraints. Among the transceivers used in the comparison, we have proposed a data rate, spectrum occupation, radio range and TH-code IR-UWB reconfigurable transceiver. This work has illustrated briefly the synchronization issue in coherent IR-UWB technique. The co-design platform could be use for work on most of classical IR-UWB transceiver architecture design.

Our future work will be oriented towards the 60 GHz UWB radio channel and the improvement of the high level platform in order to implement a WSN MAC (Medium Access Control) layer. The idea is to be able to evaluate the performance of a full multi node WSN with reconfigurable IR-UWB radio interface (PHY and MAC layer). The work presented here, is the first stage.